\documentclass[review]{elsarticle}

\usepackage{lineno,hyperref}
\modulolinenumbers[1]
\usepackage{xcolor}
\journal{Journal of \LaTeX\ Templates}

\usepackage{graphicx}
\usepackage{dcolumn}
\usepackage{bm}
\usepackage{amstext}
\usepackage{amsmath}
\usepackage{amssymb}
\usepackage{braket}

\newcommand{\vv}{\mathrm{AC}^{-} \otimes \mathrm{CR}^{-}}
\newcommand{\vt}{\mathrm{AC}^{+} \otimes \mathrm{CR}^{-}}
\newcommand{\ttag}{\mathrm{AC}^{+} \otimes \mathrm{CR}^{+}}

\newcommand{\chiv}{\chi_{\mathrm{v}}}
\newcommand{\chis}{\chi_{\mathrm{ps}}}

\begin{document}

\begin{frontmatter}

\title{Constraints on Bosonic Dark Matter with Low Threshold Germanium Detector at
  Kuo-Sheng Reactor Neutrino Laboratory}

\author[AS,BHU]{Manoj~Kumar~Singh}

\author[AS]{Lakhwinder~Singh\corref{corrauthor}}
\ead{lakhwinder@gate.sinica.edu.tw}
\author[AS,NDHU]{Mehmet~Agartioglu}
\author[AS,BHU]{Vivek~Sharma\corref{corrauthor}}
\ead{vsharma@gate.sinica.edu.tw}
\author[BHU]{Venktesh~Singh}
\author[AS]{Henry~Tsz-king~Wong}

\cortext[corrauthor]{Corresponding author}

\address[AS]{Institute of Physics, Academia Sinica, Taipei 11529, Taiwan}
\address[BHU]{Department of Physics, Institute of Science, Banaras Hindu University, 
Varanasi {221005}, India}
\address[NDHU]{Department of Physics, National Dong Hwa University, Shoufeng, Hualien 97401, Taiwan}

\begin{abstract}
  We report results from searches of pseudoscalar and vector bosonic super-weakly interacting
  massive particles (super-WIMP) in the TEXONO experiment at the Kuo-Sheng Nuclear Power Station,
  using 314.15 kg days of data from $n$-type Point-Contact Germanium detector. The super-WIMPs
  are absorbed and deposit total energy in the detector, such that the experimental signatures
  are spectral peaks corresponding to the super-WIMP mass. Measured data are compatible
  with the background model, and no significant excess of super-WIMP signals are observed. We
  derived new upper limits on couplings of electrons with the pseudoscalar and vector bosonic super-WIMPs 
  in the sub-keV mass region, assuming they are the dominant contributions to the dark matter
  density of our galaxy.

\end{abstract}


\end{frontmatter}


\section{Introduction}
\label{intro}
There is ample evidence for the existence of dark matter~(DM) that
come from numerous  astronomical and cosmological
observations at different scales. Since all the existing evidence
only probe the gravitational interaction of DM, their
non-gravitational interactions with normal matter are still
elusive. Many extensions of the standard model (SM) of particle physics predict
viable DM candidates. Weakly Interacting Massive Particles (WIMP), axions or
axion-like particles (ALPs), sterile neutrinos and millicharged particles are
leading candidates of DM. Searches for these leading candidates are in full swing,
but an experimental verification via direct, indirect detection
or production from LHC is still awaited. In the absence of a
credible positive signal, direct detection experiments have ruled out 
a substantial portion of the favored WIMP parameter space in
the GeV to TeV mass range~\cite{Xenon1T:2018, PandaX-II:2017}. 
The light-DM (LDM) has received more attention at both theoretical
and experimental front in recent years. 
Bosonic super-weakly interacting particles (denoted by $\chi$) constitute another
large category of LDM candidates
with masses at the keV-scale~\cite{superWIMP2008}. 
These bosonic $\chi$ candidates are experimentally very interesting due to their
absorption via ionization or excitation of an electron in the target-atom
of detectors. The bosonic $\chi$ would deposit the energy equivalent to their rest mass
in the detector. A photo-like peak with unexplained origin in the measured energy spectrum would be
``smoking gun'' signature of bosonic LDM. Point-Contact Germanium (PCGe)
detectors~\cite{Soma:2014} with their excellent energy resolution,
sub-keV threshold, and low intrinsic radioactivity background are the best candidates to study
LDM and other physics beyond the SM.  

The theme of this article is to report improved direct laboratory limits on $\chi$
coupling based on the data acquired by an n-type Point-Contact Germanium ($n$PCGe) detector
at the Kuo-Sheng Reactor Neutrino Laboratory (KSNL). The structure of
this report is as follows. Section~\ref{theoretical::rate} describes the interaction and expected
rate of $\chi$ at Ge target. Highlights of the experimental setup and data acquisition 
at KSNL are presented in Sec.~\ref{sec::setup}. The performance parameters of the
$n$PCGe detector are summarized in Sec.~\ref{sec::char}. The detail of event
selection of candidates and background suppression is explained in Sec~\ref{sec::dataAnalysis}.  
We present our results compared with the other representative experiments in Sec.~\ref{sec::results}. 
Finally, we conclude in  Sec.~\ref{sec::summary}.

\section{Bosonic Dark Matter}
\label{theoretical::rate}
Pseudoscalar, scalar, and vector are three generic possible candidates of non-relativistic
LDM that may have a superweak coupling with SM particles. The correct relic density of bosonic
LDM in a wider mass range could be obtained via either thermally or non-thermally misalignment
mechanism~\cite{WISpy:misalig2012, vector:misalign2011, scalar:misalign2015}.  
The bosonic pseudoscalar ($\chis$) are excellent candidates of LDM.  The
effective interaction Lagrangian of pseudoscalar-LDM particles with the electron
is given as
\begin{equation}
\mathcal{L}_{int} = \frac{g_{aee}}{2 m_{e}} (\partial_{{\mu}} a) \bar{\psi} \gamma^{\mu} \gamma^{5} \psi,
\end{equation}
where $a$ is the bosonic pseudoscalar field, $g_{aee}$ is the pseudoscalar-electron coupling constant,
$m_{e}$ is the mass of electron, $\gamma^{\mu}$ are the Dirac matrices, and $\bar{\psi}$, $\psi$
are dual Dirac spinors.  The phenomenology behind $\chis$ is
similar to nonrelativistic ALPs. The $\chis$ have coupling to atoms through the
axioelectric effect:
\begin{equation}
\chi + A  \rightarrow  A^{+} + e^{-},
\end{equation}
which is analogous to the photoelectric effect with the absorption
of $\chis$ instead of a photon~\cite{axioeffect:1986}. The only difference between
photoelectric effect and axioelectric effect is the wave function of absorbed photon 
 and $\chis$. The wave function of absorbed photon contains a space-dependent factor
 exp({$i\bf{kr}$}), which is replaced by exp(${i m_{\mathrm{ps}} \bf{v_{\mathrm{\chi}}r}}$)
 in case of massive $\chis$~\cite{superWIMP2008}. Here $m_{\mathrm{ps}}$ and $v_{\mathrm{\chi}}$
 are the mass and velocity of the incoming $\chis$  particle, respectively.
 The absorption cross section $\sigma_{abs}$ (axio-electric effect)
for $\chis$ can be written as
\begin{eqnarray}
\sigma_{abs} \simeq \frac{3m_{\mathrm{ps}}^{2}}{4\pi\alpha f^{2}_{a} \beta} {\sigma_{pe}(w= m_{\mathrm{ps}})},
\end{eqnarray}
where $\sigma_{pe}$ is the photoelectric cross section with the photon energy $\omega$ replaced by
the mass of $\chis$ ($m_{\mathrm{ps}}$), $f_{a}$ = 2$m_{e}/g_{aee}$ is the dimensionless
coupling strength of $\chis$ to SM particles and $\beta \equiv v_{{\chi}}/c$.
The absorption cross section of $\chis$ is directly proportional to
$m_{\mathrm{ps}}^{2}$.

The best motivated model for bosonic vector dark matter ($\chi_{\mathrm{v}}$) is kinetic mixing model,
in which an extra U(1)$_{D}$ gauge group is introduced into SM gauge group. The  kinetic mixing with the hypercharge
field strength is responsible for the interaction between the ordinary matter and $\chi_{\mathrm{v}}$. 
The interaction Lagrangian density for $\chi_{\mathrm{v}}$ with electrons after breaking of U(1)$_{D}$
can be expressed as

\begin{equation}
\mathcal{L}_{int} = {\kappa}{e} V_{\mu} \bar{\psi} \gamma^{\mu} \psi,
\end{equation}
where  $V_{\mu}$ is the bosonic vector field couples to the electromagnetic current,
$e$ is the electronic charge, $\kappa$ is the vector
hypercharge. The product $e'$ = $e\kappa$ is analogous to electromagnetic
coupling such that $\alpha'$ = (e$\kappa$)$^{2}$/4$\pi$ is similar to a
vector-electric fine-structure constant. The Compton wavelength of $\chi_{\mathrm{v}}$
is much larger than the linear dimension of the atom due to nonrelativistic behavior.
This allow the multipole expansion in the interaction. The contribution of E1-transition dominates over
other multipoles~\cite{DP-astro:2015}, which makes the matrix element of photon absorption proportional
to the matrix element of $\chiv$ absorption. The absorption cross section of
$\chiv$ ($\sigma_{abs}$) can be expressed in the photoelectric effect
with the replacement of photon energy $\omega$  by the mass m$_{\mathrm{v}}$ of $\chiv$
and the coupling constant is scaled appropriately as \cite{superWIMP2008}
\begin{eqnarray}
\frac{\sigma_{abs}}{\sigma_{pe}(\omega = m_{\mathrm{v}}) } \simeq \frac{\alpha'}{\alpha} \frac{1}{\beta}.
\end{eqnarray}
The absorption cross section of $\chiv$ is nearly independent of the DM mass.

The key difference between $\chis$ and $\chiv$ LDM
is their decay modes to photons. The two-photon decay channel is strictly forbidden for 
vector boson, whereas decay into three photons is allowed at loop level~\cite{DP-astro:2015}.
Pseudoscalar bosons are axion-like with two-photon decay as a dominant channel.  
The mass of a scalar particle receives quadratically radiative contributions from its interactions
with other SM  particles. The keV-scale bosonic scalar ($\chi_{\mathrm{s}}$) requires highly model
dependent parameters in order to make it viable candidates. The phenomenology of $\chi_{\mathrm{s}}$ is
also quite similar to the $\chis$ in many ways~\cite{superWIMP2008}. Experimental searches thus mainly
focus on $\chis$ and $\chiv$.

Assuming the  $\chi$ ($\chis$ or $\chiv$) constitutes all of the galactic DM with their local
density $\rho_{\chi}$ = 0.3 GeV/cm$^{3}$, the total average flux of
$\chi$ ($\Phi_{\chi}= \rho_{\chi} v_{\chi}/m_{\chi}$) is given by 
\begin{eqnarray}
   \Phi_{\chi} = \frac{9.0 \times 10^{15}}{m_{\chi}} \times \beta ~~~\mathrm{cm}^{-2} \mathrm{s}^{-1},
\end{eqnarray}
where m$_{\chi}$ is the mass of $\chi$  in keV/c$^{2}$,
$v_{\chi}$ is the average velocity of $\chi$ relative to the Earth; 
$\beta \equiv v_{{\chi}}/c \; \sim$ 10$^{-3}$ implies that the energy deposition by bosonic $\chi$
would be E = $\sqrt{(m_{\chi}^{2} + \gamma^{2} m_{\chi}^{2} \beta^{2})}\sim m_{\chi}$.

The theoretically expected interaction rate of $\chi$ in a direct detection experiment 
can be expressed as
\begin{eqnarray}
  R_{\chi} =  \rho_{d} \sigma_{abs} \Phi_{\chi},
  \label{EQ::expected-rate}
\end{eqnarray}
where $\rho_{d}$ = N$_{A}$/A is the atomic number density per unit target mass of detector and N$_{A}$ is
the Avogadro's number. $\sigma_{abs}$ is the absorption cross section of $\chi$. 
\begin{figure}
  \centering
  \includegraphics[width=9.0cm]{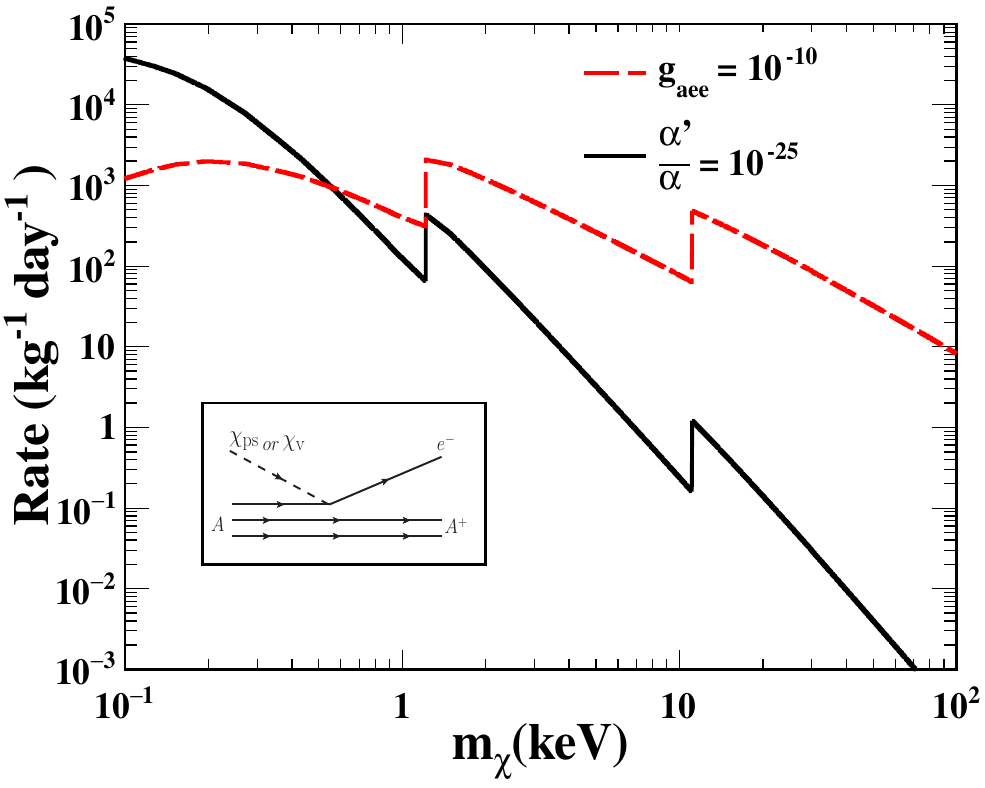}
  \caption{The expected pseudoscalar and vector interaction rates in Ge-detector, for fixed values of
    $g_{aee}$ and $\alpha'$/$\alpha$, as a function of $m_{\chi}$. 
    The axioelectric interaction is illustrated schematically in the inset.}
\label{fig::BDM-rate}
\end{figure}
The interaction rate of $\chis$ in the units of kg$^{-1}$day$^{-1}$ can be calculated by using
the Eq.~\ref{EQ::expected-rate}. Accordingly, the expected rate becomes
\begin{eqnarray}
  R_{\mathrm{ps}} \simeq \frac{1.2 \times 10^{19}}{A} g_{aee}^{2} \frac{m_{\mathrm{ps}}}{[\mathrm{keV}]} \frac{\sigma_{pe}}{[\mathrm{barns}]} \mathrm{kg}^{-1}\mathrm{day}^{-1},
  \label{eq::scalar-rate}
\end{eqnarray}
where A is the atomic mass of target materials. Similarly, the expected count rate of
$\chiv$ in the direct detection experiment can be expressed as
\begin{eqnarray}
 R_{\mathrm{v}} \simeq \frac{4 \times 10^{23}}{A} \frac{\alpha'}{\alpha} \frac{[\mathrm{keV}]}{m_{\mathrm{v}}} \frac{\sigma_{pe}}{[\mathrm{barns}]}~ \mathrm{kg}^{-1}\mathrm{day}^{-1}.
\label{eq::vector-rate}
\end{eqnarray}

Point-Contact Germanium detectors~\cite{Barbeau:pge2007, Soma:2014} are ideal
devices for the searches of $\chis$ and $\chiv$, due to their low energy
threshold which expand the sensitivities to low $m_{\chi}$ and excellent energy resolution 
allows spectral peaks to be resolved. The interaction rates of $\chis$ and
$\chiv$ with Ge-detector follow Eqs.~(\ref{eq::scalar-rate}) and~(\ref{eq::vector-rate}),
respectively, are illustrated in Fig.~\ref{fig::BDM-rate} for specific
choices of the coupling constants. The characteristic features in the calculated
spectrum at energies 11.10~keV~(K-shell) and 1.41~keV~(L-shell) correspond to absorption
edges of the photoelectric effect in the germanium atom.  These features as peaks or edges 
at the specific binding energies significantly enhance the sensitivity of the experiment
compared to the case of a continuum curve.

\section{Experimental Setup}
\label{sec::setup}
The TEXONO collaboration~\cite{Wong:2018} aims to progressively
improving the sensitivities towards electromagnetic properties of
neutrino~\cite{ntuEM:2015,ntuSterile:2016,txnPRD:nmm2010} as well as
WIMP~\cite{txnPRL:DM2013,txnPRD:DM2009}, axions~\cite{txnPRD:ax2007} and
physics searches beyond the SM at KSNL. The experimental setup is placed at
the first floor of the seven-storey reactor building at depth of $\sim$12 m
below the sea level. The concrete of the reactor building provides
30 meters of water equivalent overburden, which eliminates most of
the hadronic components of cosmic rays. A multipurpose 4$\pi$ passive
shielding house with ``inner target'' detector volume of 100 cm $\times$ 80
cm $\times$ 75 cm consists of, from the inside out, 5 cm of oxygen-free
high-conductivity copper, 25 cm of boron-loaded polyethylene,
5 cm of steel, 15 cm of lead. The shielding house is further
mounted with cosmic ray (CR) veto scintillator panels. 
This 50-ton passive shielding house provides  significant
attenuation to the ambient neutron and $\gamma$ backgrounds. 
The Ge-detector surrounded by 38.3 kg well-shaped NaI(Tl)
anti-Compton (AC) detector is placed inside the inner target volume. 
A detailed description of shielding and the detector development at
KSNL can be found in Refs.~\cite{Wong:2018, txnPRD:nmm2010, Deniz:2010}.

\begin{figure*} 
  \includegraphics[width=\textwidth]{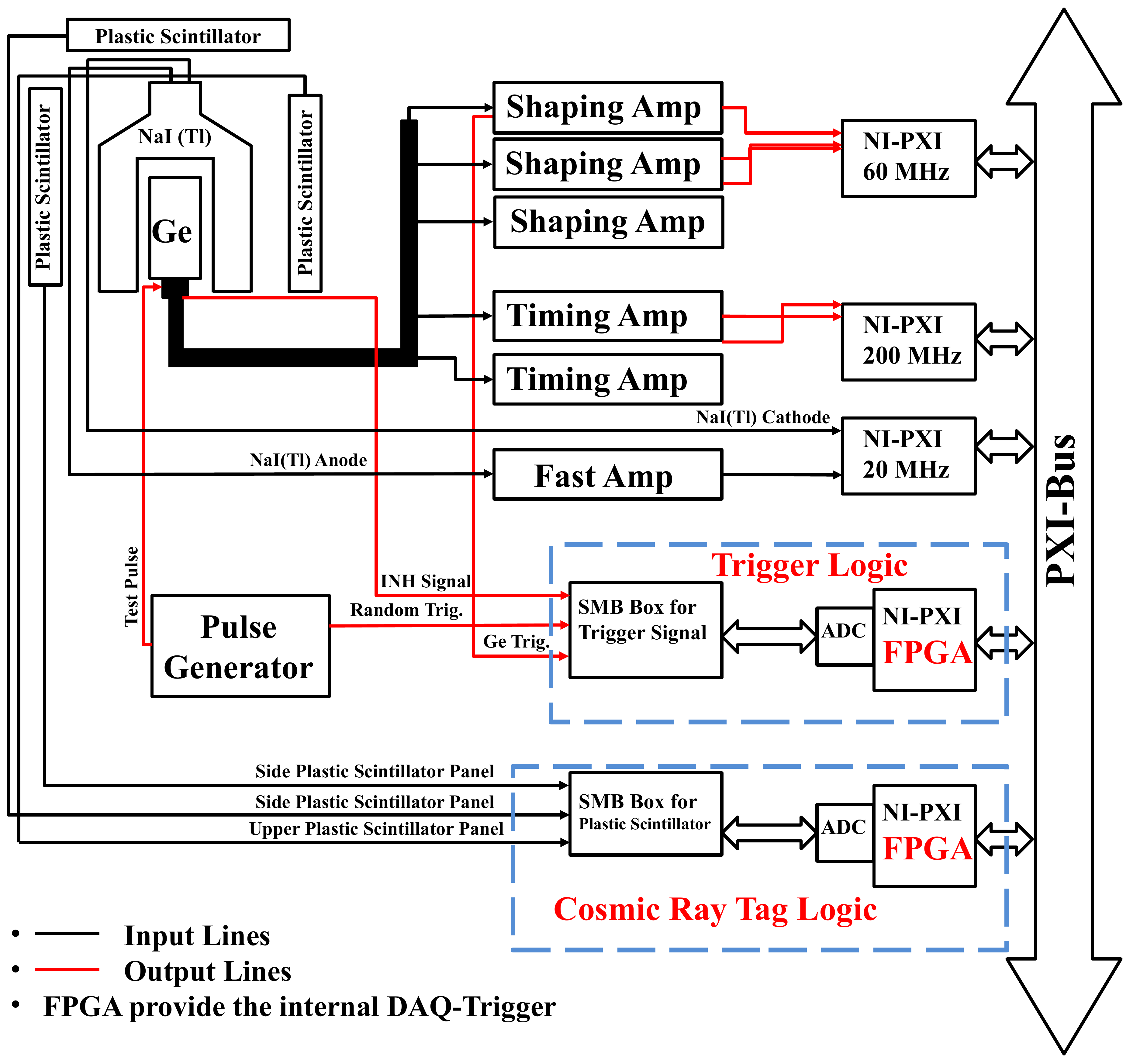} 
  \caption{Schematic diagram of the data acquisition system for TEXONO experiment at KSNL.}
  \label{fig::DAQ-Diagram}
\end{figure*}

The data acquisition (DAQ) system in operation at KSNL is a hybrid design 
of digital and analog electronics. The schematic block-diagram of electronics 
and DAQ system is illustrated in Fig.~\ref{fig::DAQ-Diagram}. The preamplifier of
PCGe is custom designed with 5 copies of a signal. Two copies of Ge-preamplifier signal are shaped
with Canberra 2026 shaping amplifier (SA) at 6~$\mu s$ and 12~$\mu s$ with
same amplification factors. The distinct behavior of self-trigger noise at
different shaping time helps to suppress noise events near the threshold.
One copy of Ge-preamplifier is processed with SA 
at minimum gain to study the high energy background behavior. 
The timing amplifier output (TA) preserves the rise-time information. Therefore,
two copies of Ge-preamplifier signal are processed with Canberra 2111 TA at ``out'' 
differentiate along with 10 $ns$ integration control setting with different gains
to cover a wide energy range from sub-keV to few hundreds keV.  

\begin{figure} 
  \includegraphics[width=\textwidth]{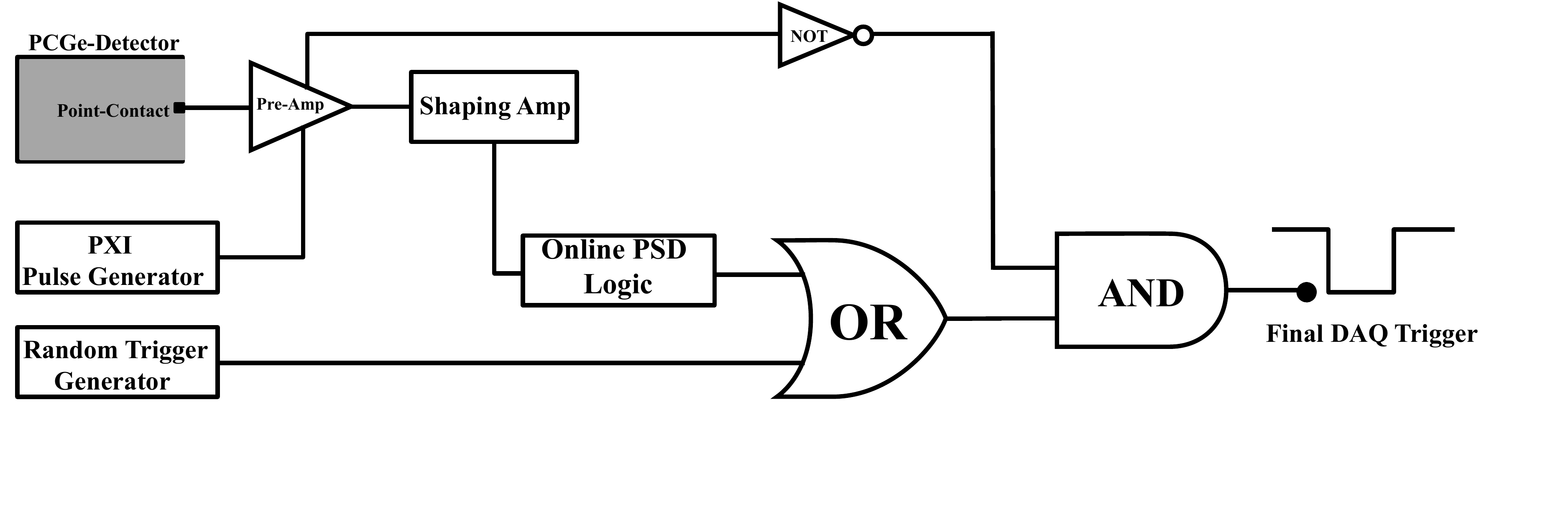}
  \caption{Schematic diagram of data acquisition trigger. }
  \label{fig::crktdig}
\end{figure}

\begin{figure*} 
  \centering
  \begin{minipage}[t]{0.48\textwidth}
    \includegraphics[width=\textwidth]{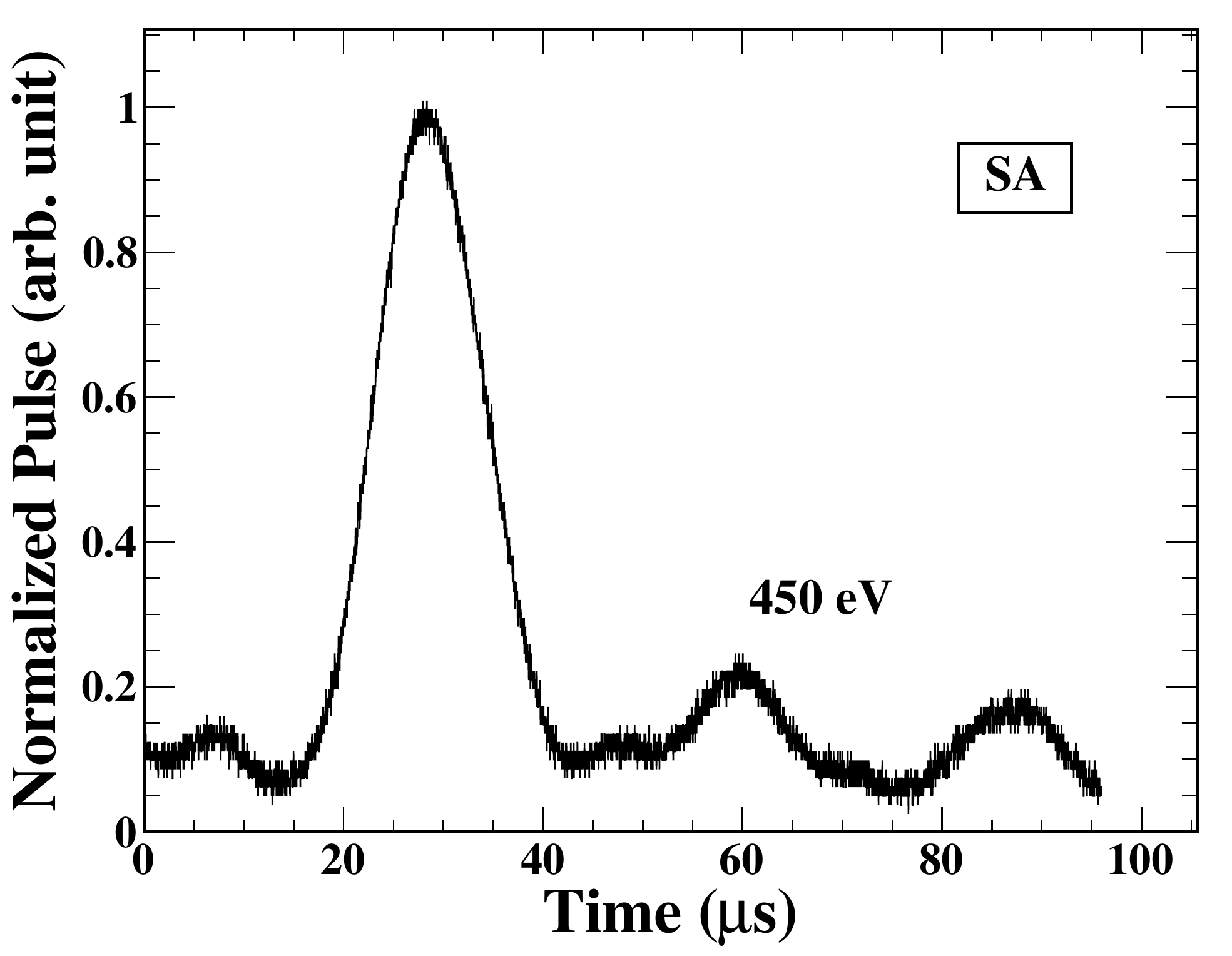}
    \caption{A Typical pulse shape at 450 eV from SA at 6 $\mu s$ output. }
    \label{fig::pulse60m}
  \end{minipage}
  \hfill
  \begin{minipage}[t]{0.48\textwidth}
    \includegraphics[width=\textwidth]{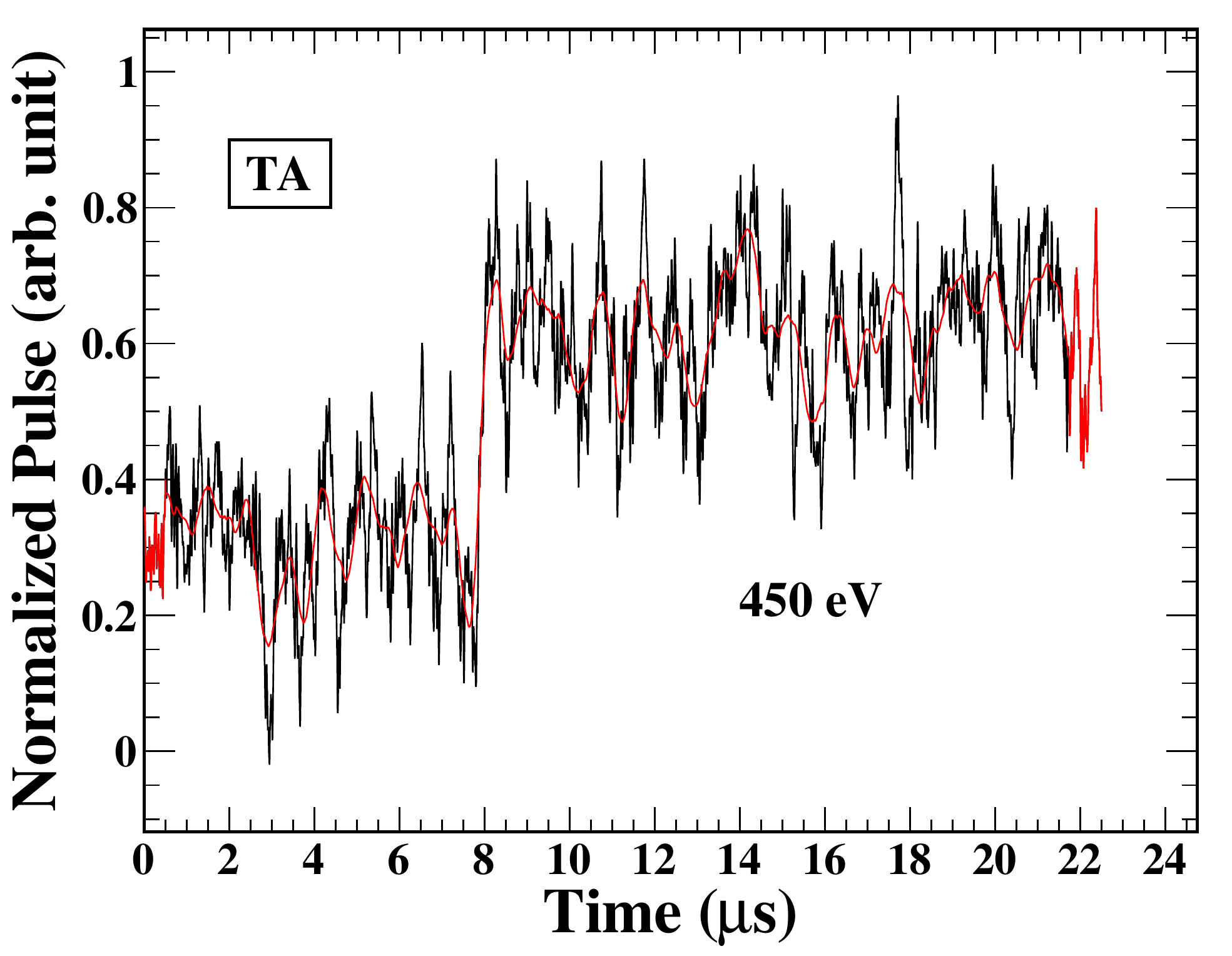}
    \caption{A Typical pulse shape at 450 eV from TA output with
      raw (black) and smoothed (red) pulse.}
    \label{fig::pulse200m}
  \end{minipage}
  \hfill
\end{figure*}

The KSNL readout system is built around National Instruments 
(NI) PXIe-1065 chassis with an embedded PXI-8108 controller and extended 
by NI PXI-5105 analog-to-digital converter (ADC), NI PXI-5124 ADC, and
NI PXIe-7961R FlexRIO Field Programmable Gate Arrays (FPGA) cards. 
The NI PXI-5105 ADC with 12-bit resolution, 60 MHz bandwidth and 60 MS/s~{(MS:~Mega~sample)} real-time sample
rate is  being used to digitize the signal from shaping amplifiers. 
The NI PXI-5124 ADC with 12-bit resolution, 200 MS/s real-time sample rate and
150 MHz bandwidth records the signal from timing amplifiers. 
The NI PXIe-7961R FlexRIO FPGA module composed with high-speed 16-channel,
14-bit and 50 MS/s NI 5751 digitizer-adapter is the heart of the
DAQ system. These FPGA-modules provide the user-defined logic to specify
the trigger condition for CR veto scintillator panels and Ge-detector.
It also helps to reject some microphonic noise online, based on distorted pulse shape.
The signal from the DC-coupled preamplifier is discriminated after being
shaped with SA at 6~$\mu s$ served as a Ge-trigger. The random-trigger is generated
by a function generator at {the} rate of 0.1 Hz and recorded randomly due to OR-{gate} logic 
with random Ge-trigger. These random-trigger events help to measure the 
pedestal fluctuation and dead time of the DAQ system. 
The inhibit output from the preamplifier is used to veto all triggers
generated during the reset of the preamplifier, including random-trigger.
A schematic diagram of DAQ trigger scheme at  KSNL is shown in Fig.~\ref{fig::crktdig}.
A typical low energy signal from SA and TA is depicted in Fig.~\ref{fig::pulse60m} and
Fig.~\ref{fig::pulse200m}, respectively.
The signal of AC NaI-detector and CR veto panel are also recorded along with Ge-trigger
to understand the cosmic induced background, and filtered out in the offline analysis.
The known amplitude pulses are generated from the precision wave {function} generator 
and injected into the test port of the preamplifier to study 
the trigger efficiency and electronic behavior of the DAQ system. 
The DAQ software is developed in NI LabVIEW programming 
language which provides a user-friendly environment.
To monitor the data quality, the DAQ software generates
cumulative diagnostic plots for immediate inspection.

\section{Characterization of N-type PCGe-Detector}
\label{sec::char}
\begin{figure*} 
  \centering
  \begin{minipage}[t]{0.48\textwidth}
    \includegraphics[width=\textwidth]{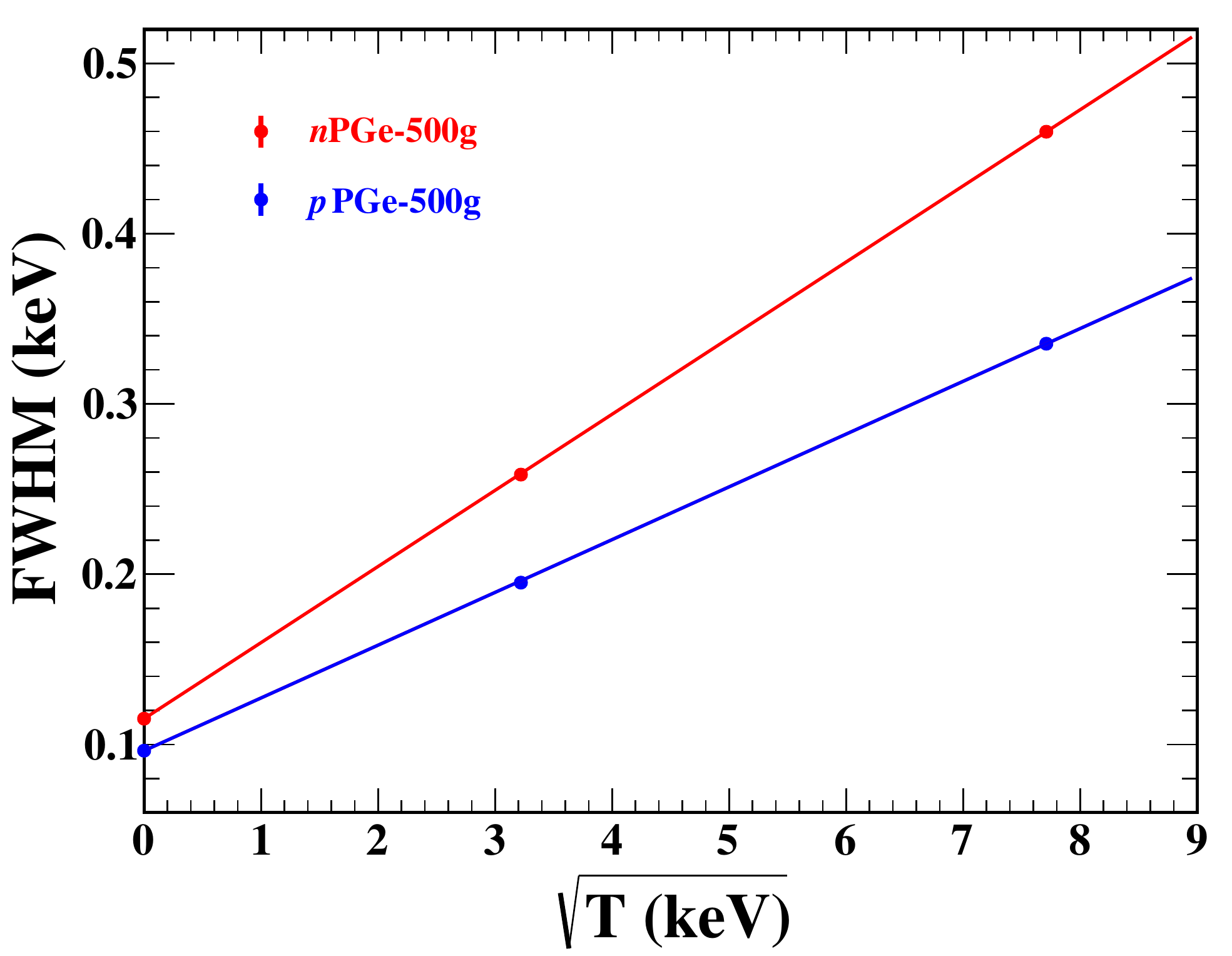}
    \caption{Comparison of energy resolution of $n$PCGe and $p$PCGe at pedestal fluctuation (0 keV), Ga K-Shell
    X-ray (10.37 keV) and $^{241}$Am $\gamma$-ray (59.5 keV).}
    \label{fig::reso}
  \end{minipage}
  \hfill
  \begin{minipage}[t]{0.48\textwidth}
    \includegraphics[width=\textwidth]{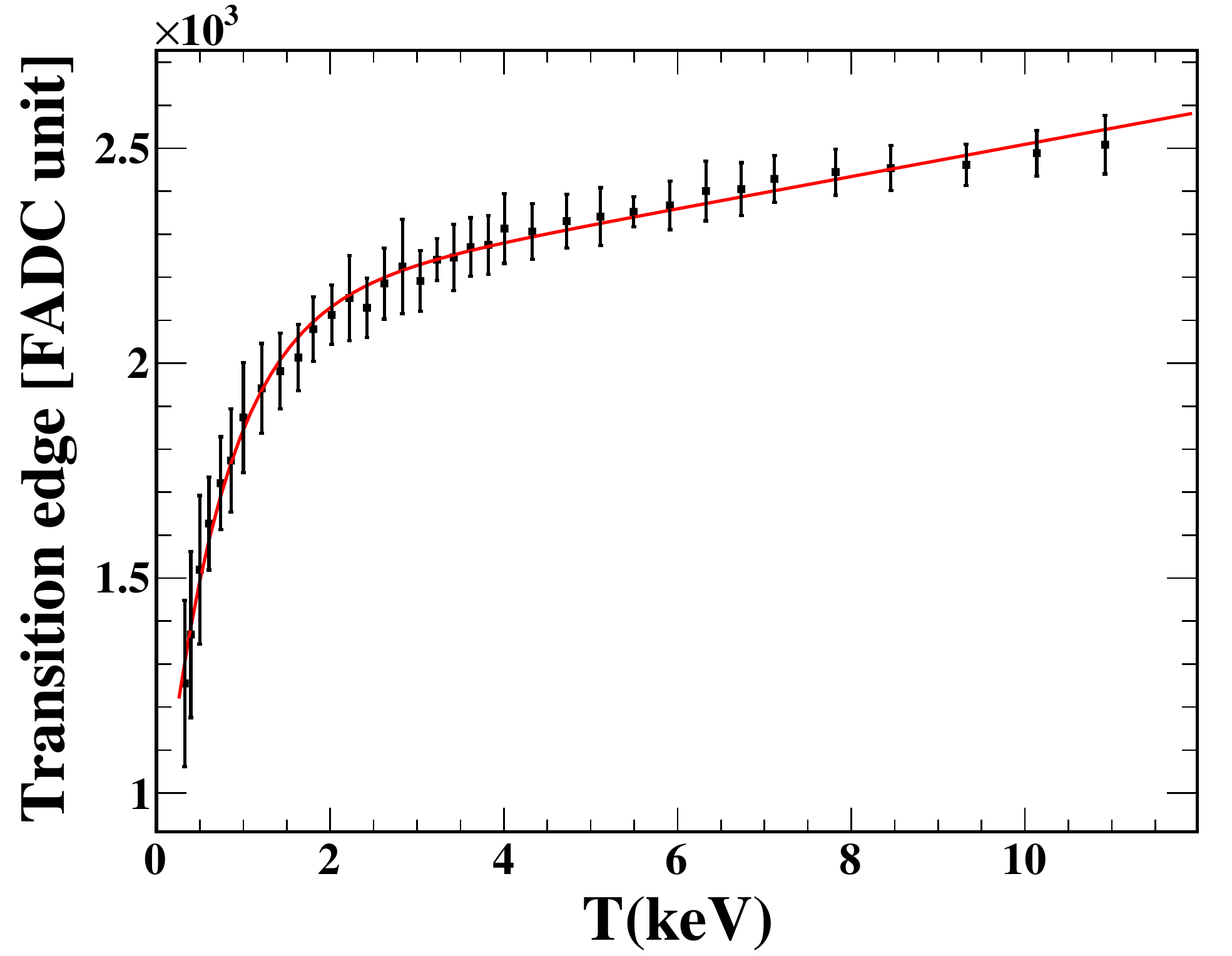}
    \caption{The transition edge of TA-pulses as a function of
      energy with best fit function.}
    \label{fig::trans-edge}
  \end{minipage}
  \hfill
\end{figure*}

The energy calibration from sub-keV to 14 keV is 
achieved by the combination of precise-pulser measurement and X-ray peaks 
(10.37 keV and 1.29 keV), resulting from electron capture of Ga daughter induced 
by cosmogenic  {$^{68,71}$Ge}-isotope. The maximal amplitude and
integrated area of the pulse are both energy related parameter{s}. The maximal  
amplitude is adopted to estimate the energy of an event for its better
energy resolution~\cite{Soma:2014, IJP:BHU2018}.
The energy resolution of $n$PCGe is comparable to p-type Point-Contact 
($p$PCGe) in low energy region, but degraded in high energy region.
The energy resolution of $n$PCGe may be enhanced by
increasing the detector operating temperature, selecting the material
with lower electron trap concentrations and higher gradients impurity concentrations
to generate higher drift field \cite{Luke:ppc1989, Barbeau:pge2007}.
Figure~\ref{fig::reso} illustrates the variation of full width at half maximum (FWHM) 
as a function of energy for $n$PCGe and $p$PCGe detectors. The FWHM of $n$PCGe degrades as
energy deposition increases. The performance parameters of same modular mass
$n$PCGe and $p$PCGe detectors are summarized in Table~\ref{pntable}.

\begin{table}
  \caption{Comparison between $n$PCGe and $p$PCGe-detector on basis of various performance parameters.
    The pulse maximal amplitude is adopted as energy estimator in sub-keV energy region.}
  \begin{tabular}{lccc}
    \hline
      Item                                         & $p$PCGe       & $n$PCGe      \\
      \hline
      Modular Mass (g)                             &  500          & 500          \\
      RESET Amplitude (V)                          &  6.8          & 6.8          \\
      RESET Time Interval (ms)                     &  $\sim$160    & $\sim$ 170   \\ 
      Pedestal Noise                               &               &              \\
      ~~~ Amplitude RMS $\sigma_{Noise}$ (eV)       &  41            & 49           \\
      Pulser Width                                 &               &              \\
      ~~~FWHM (eV)                                 &  110          & 133          \\
      ~~~RMS (eV)                                  &  47           & 52           \\
      X-Ray Line Width                             &  Ga-K         & Ga-K         \\
      ~~~RMS (eV)                                  &  87           & 104          \\
      Electronic Noise-Edge for Raw Spectra (eV )  &  228          & 285          \\
      \hline      
    \end{tabular}
  \label{pntable}
\end{table}

\begin{figure*}
  \centering
  \begin{minipage}[t]{0.48\textwidth}
    \includegraphics[width=\textwidth]{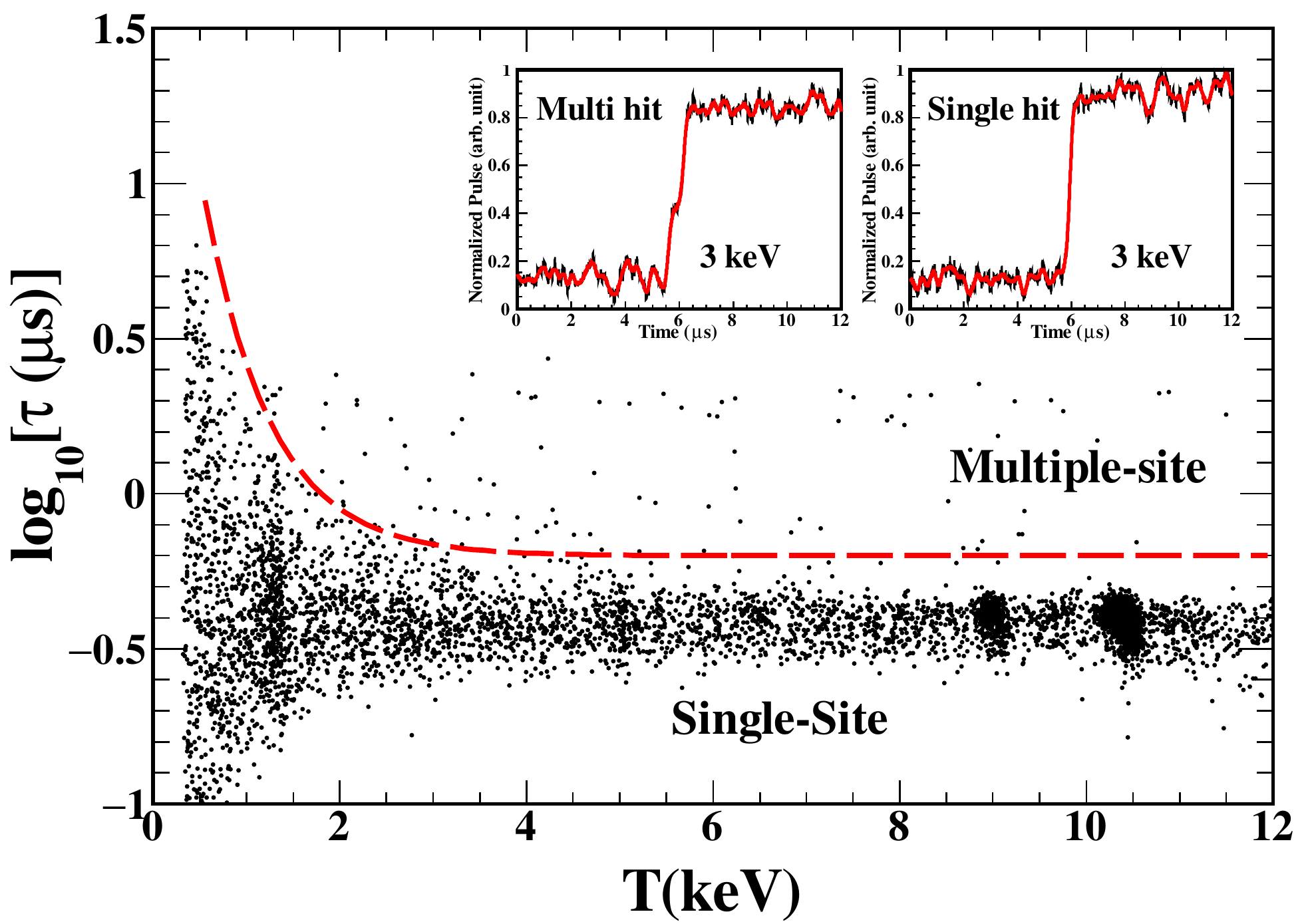}
    \caption{The scatter plot between $\tau$ and T 
      for the $\vv$-tag events shows absence of 
      anomalous surface events. Typical single- and multiple-site events are shown
      in inset, multiple-site events are characterized by kinks in rise-time profiles.}
    \label{fig::taudist}
  \end{minipage}
  \hfill
  \begin{minipage}[t]{0.48\textwidth}
    \includegraphics[width=\textwidth]{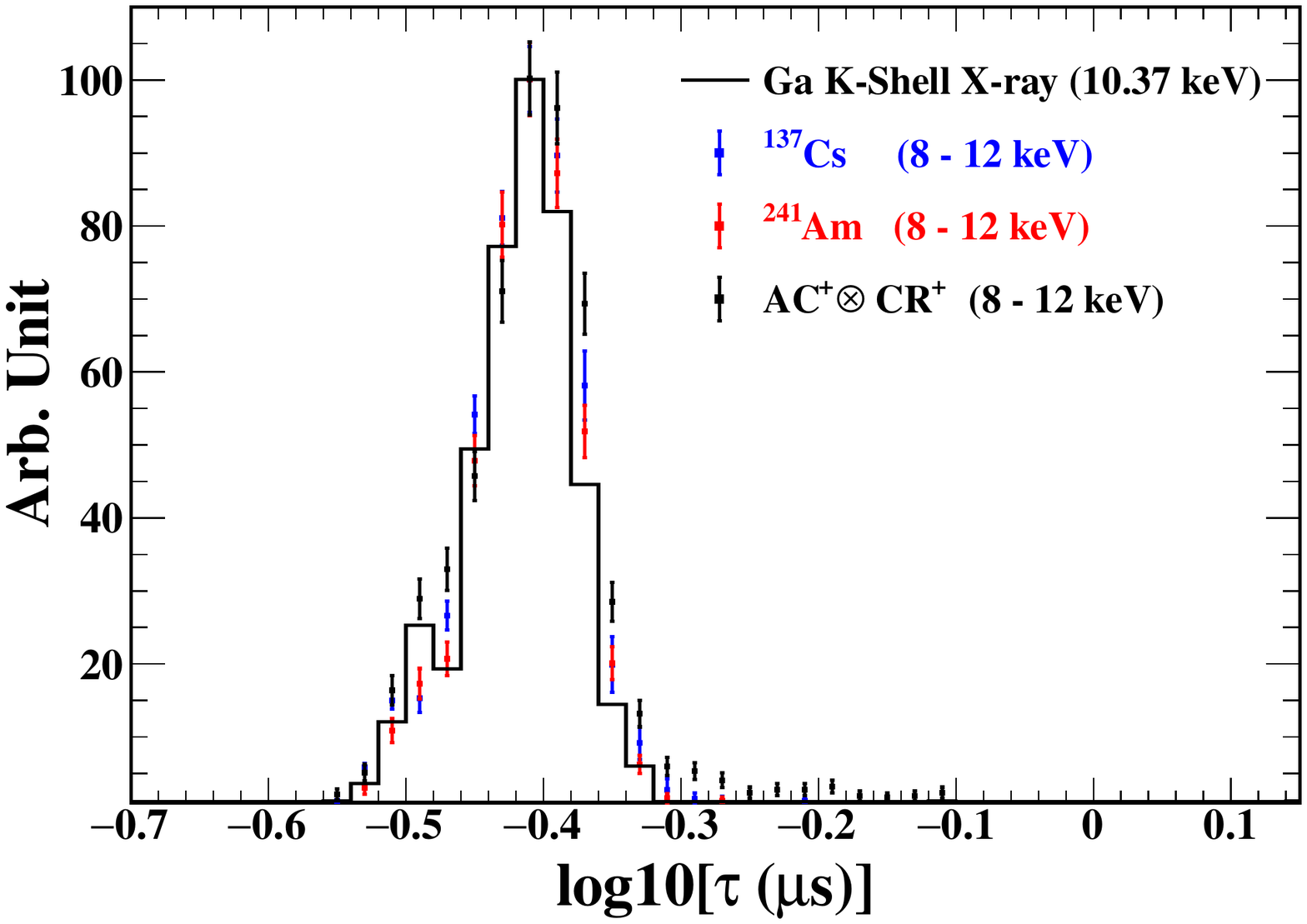}
    \caption{ The $\tau$-distribution of Ga K-Shell X-ray (10.37 keV), 
      comparing the candidate events with those of $^{241}$Am, $^{137}$Cs between 
      8 to 12 keV energy region.}
    \label{fig::npctau}
  \end{minipage}
  \hfill
\end{figure*}

Experimental sensitivity for physics beyond the SM is not only 
limited by signal detection threshold of the Ge-detectors, but also 
due to the degradation of background rejection capability at 
sub-keV recoil energy region. The anomalous surface events in 
$p$PCGe arise from a few millimeters thick transition layer 
with a weak electric field. The poor charge collection in this 
region reduces the ionization yield of the events and enhances the 
background at sub-keV energy region. The p$^{+}$ outer surface 
conductive contact of $n$PCGe is fabricated with {the} boron implantation technique.
The thickness of p$^{+}$ contact on the outer surface of the crystal
is the order of $\mu$m. The anomalous surface effect in $n$PCGe is studied
by the rise time of events. In order to measure the rise time of events,
the output of TA is fitted with the hyperbolic tangent function
\begin{eqnarray}
f(t) = \frac{A_{0}}{2}{\mathrm{tanh}}[(t-t_{0})s_{0}] + P_{0},
\label{eq-rtc}
\end{eqnarray}
where A$_{0}$, P$_{0}$, t$_{0}$ and $s_{0}$ are, respectively, 
the amplitude, pedestal offset, transition edge and slope of 
the TA-pulse. The values of A$_{0}$ and P$_{0}$ are evaluated 
from the TA-pulses through the difference and mid level of 
asymptotic levels, respectively. The transition edge of 
TA-pulses as a function of energy is pre-determined as 
shown in Fig.~\ref{fig::trans-edge} and provides 
constraint on t$_{0}$. The bending attribute between 
energy and transition edge of TA-pulse at low energies is due to
the difference in shaping times of SA-trigger and TA pulse. 
The rise time ($\tau$) of TA-pulse is characterized by the slope parameter $s_{0}$ and 
rise time from 5\% to 95\% is evaluated by $\tau~=~\mathrm{log}(19)/{s_{0}}$.
The distribution of $\tau$ as a function of energy is 
depicted in Fig.~\ref{fig::taudist}, which demonstrates 
the absence of anomalous surface events.
The scattered events above the curve (red) in Fig.~\ref{fig::taudist} are identified
as multiple sites events. Typical single- and multiple-site events at low energy are 
illustrated in the inset of Fig.~\ref{fig::taudist}.
The $\tau$-distribution of events from different known origins 
can verify the uniform timing response over the entire detector
fiducial volume. Figure~\ref{fig::npctau} illustrates the uniform timing response
for external $\gamma$-ray from  $^{241}$Am, $^{137}$Cs and internal 
X-rays at 10.37 keV from $^{68,71}$Ge{,} which decay via electron capture.

\section{Data Analysis}
\label{sec::dataAnalysis}
Data taken with the $n$PCGe detector have been adopted to place constraints on
milli-charged neutrino~\cite{txn-mQ:2014} and dark matter cosmic ray~\cite{txn-DCR:2018},
as well as heavy sterile neutrinos  as dark matter~\cite{ntuSterile:2016}.    
The data presented in this article were acquired over 628.3 live-days (whole exposure)
with 500g fiducial mass of $n$PCGe-detector. The event information from
AC and CR-detectors is recorded  along with Ge-trigger. Thus,
every Ge-trigger is categorized  by AC$^{+(-)} \otimes CR^{+(-)}$,
where the superscript +(-) denotes coincidence (anti-coincidence).
The  $\ttag$-tag events are mostly due to energy deposited by
radiations in the Ge-detector. These double tagged events serve as the
reference samples to differentiate signal from noise.
\begin{figure*} 
  \centering
  \begin{minipage}[t]{0.48\textwidth}
    \includegraphics[width=\textwidth]{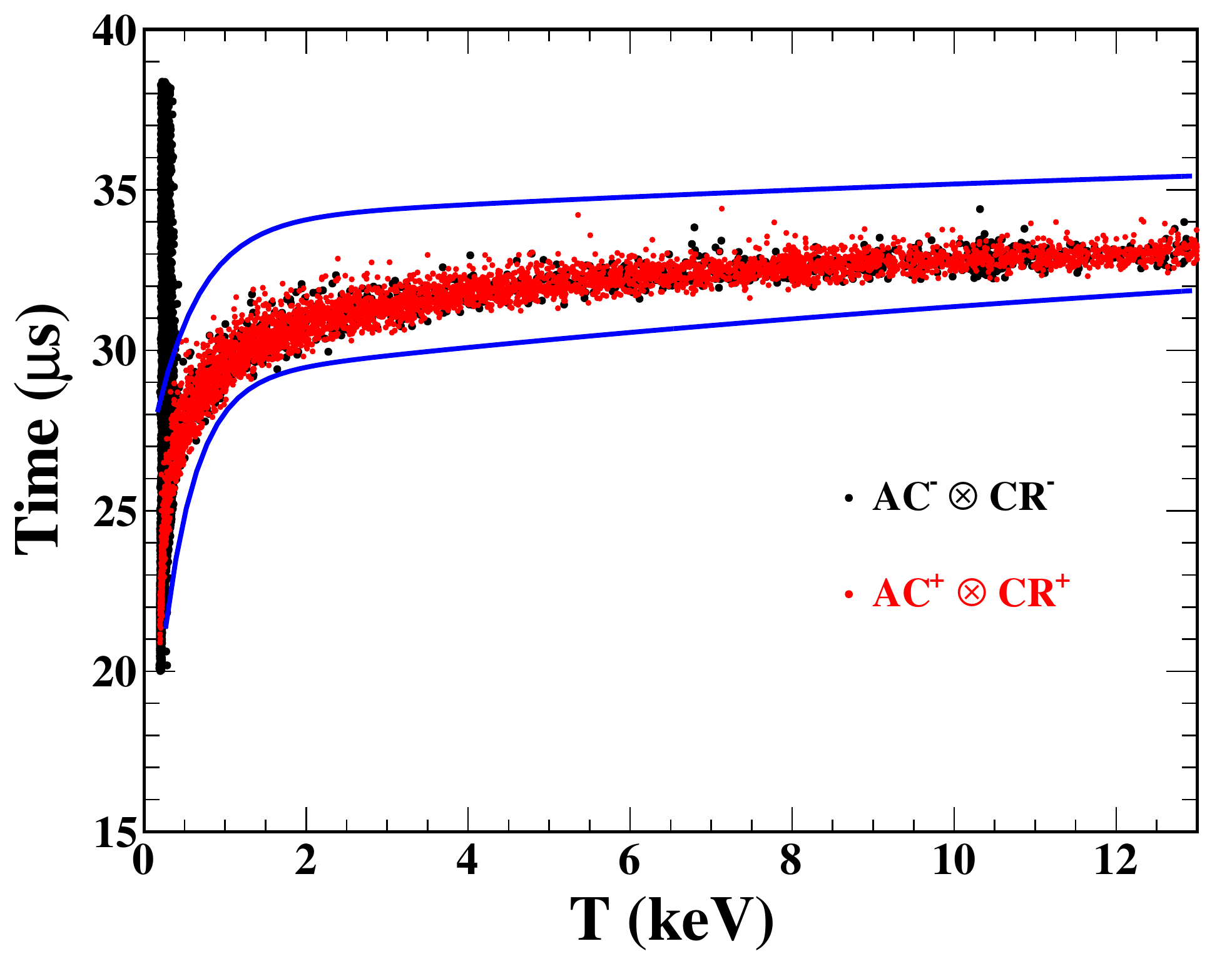}
    \caption{Scatter plot of the correlation between maximum time bin of pulse and corresponding energy.}
    \label{fig::timebincut}
  \end{minipage}
  \hfill
  \begin{minipage}[t]{0.48\textwidth}
  \includegraphics[width=\textwidth]{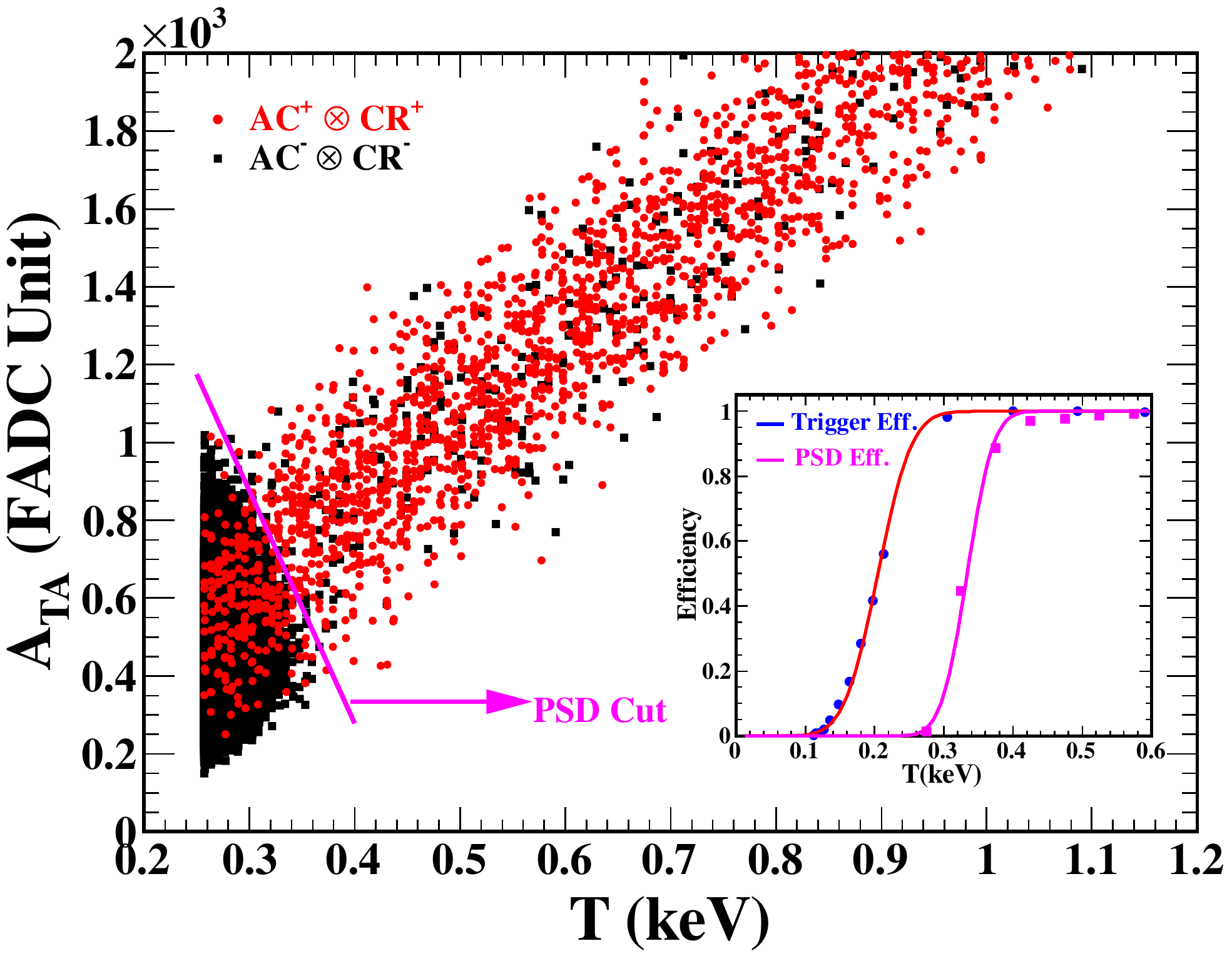}
  \caption{ Scatter plot between amplitude of TA-pulse and measured energy T 
    to illustrate the PSD selection. The inset illustrates the trigger efficiency as {a} function of energy,
    along with total PSD efficiency.}
  \label{fig::edgecut}
  \end{minipage}
  \hfill
\end{figure*}

The $\vv$-tag events are uncorrelated with AC and CR active shielding detectors
and may be {the} candidate events of $\chi$, neutrino, internal cosmogenically 
induced background and self-trigger (electronic noise)
near the noise-edge.  The threshold of Ge-detector is
restricted by self-trigger and microphonic noise. In order to avoid the contamination of 
noise near the detector threshold, it is necessary to define 
a precise signal acceptance. Therefore, pulse shape 
discrimination (PSD) technique is adopted to remove a majority of microphonic pulses
and define noise edge for analysis. The shaped waveform of Ge-detector is characterized
by several parameters. Lowest-level, pedestal (average value of pre-trigger baseline) and
maximum time bin are used to discard the most of {the} noise{,} which leaks to high energy bins.  
The maximum time bin of the good pulses lies within a time interval after the trigger. Thus, 
the amplitude and the maximum time bin of the signal pulse 
exhibit the strong correlation as shown in Fig.~\ref{fig::timebincut}, and all
events outside the correlated band are rejected. In order to define 
the noise edge between self-trigger and physics events, 
the energy of event and amplitude of TA-pulse (A$_{TA}$)
after process with {the} trapezium-filter provide one of the best possible  
parameter space. Figure~\ref{fig::edgecut} shows the distribution of
$\vv$ and $\ttag$ with the noise-edge cut.
The fraction of $\ttag$-tag events survives as
an efficiency of the PSD cuts. The trigger efficiency measured by survival
fraction of precise pulser at given 
trigger threshold. The trigger efficiency and combined PSD efficiency is 
100\% above 300 eV and 400 eV, respectively, as shown in the inset of
Fig.~\ref{fig::edgecut}.  

\begin{figure*} 
  \centering
  \begin{minipage}[t]{0.48\textwidth}
    \includegraphics[width=\textwidth]{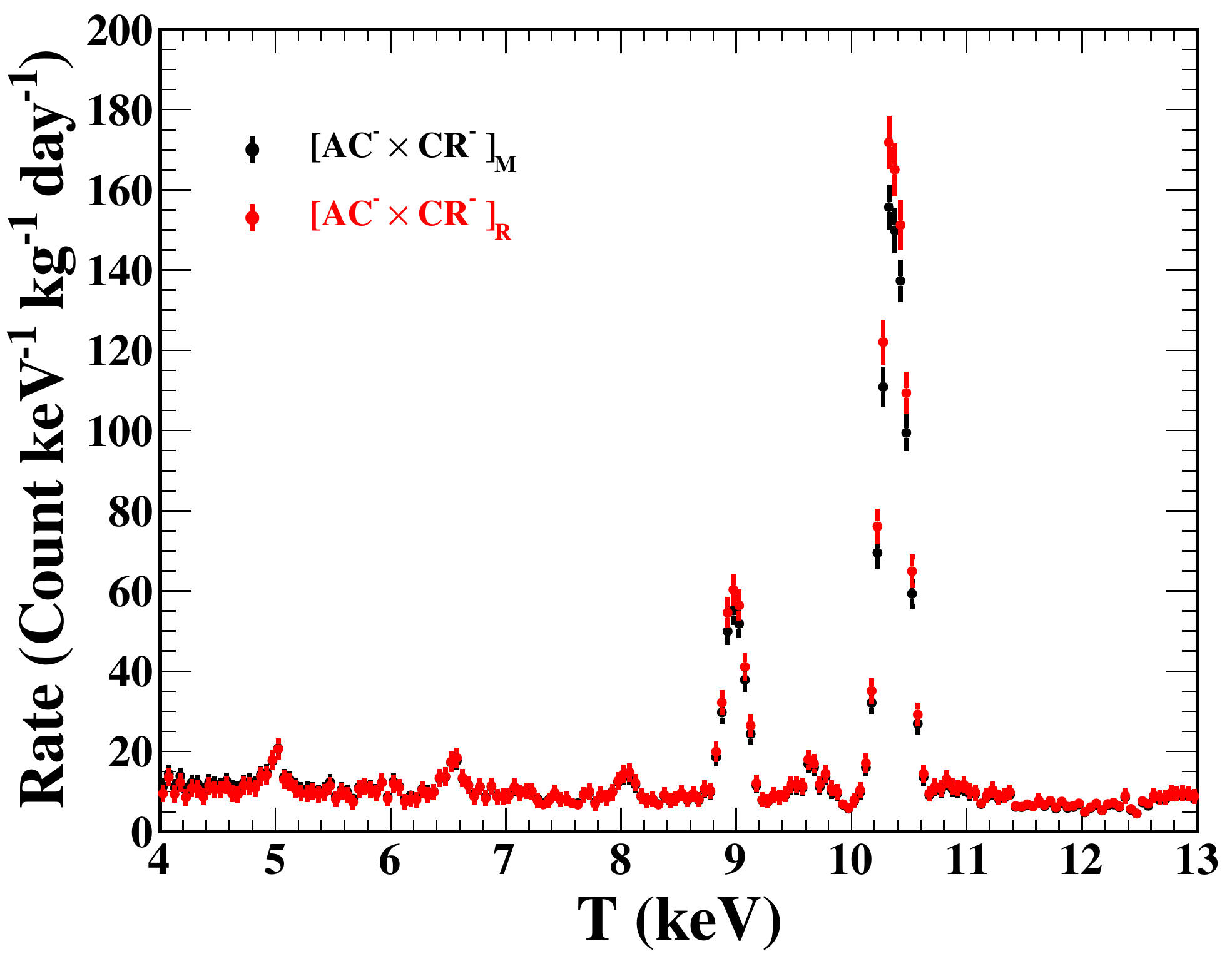}
    \caption{Measured energy spectrum of data sample, $\vv$
      before and after cosmic ray correction.}
    \label{fig::vvcorr}
  \end{minipage}
  \hfill
  \begin{minipage}[t]{0.48\textwidth}
    \includegraphics[width=\textwidth]{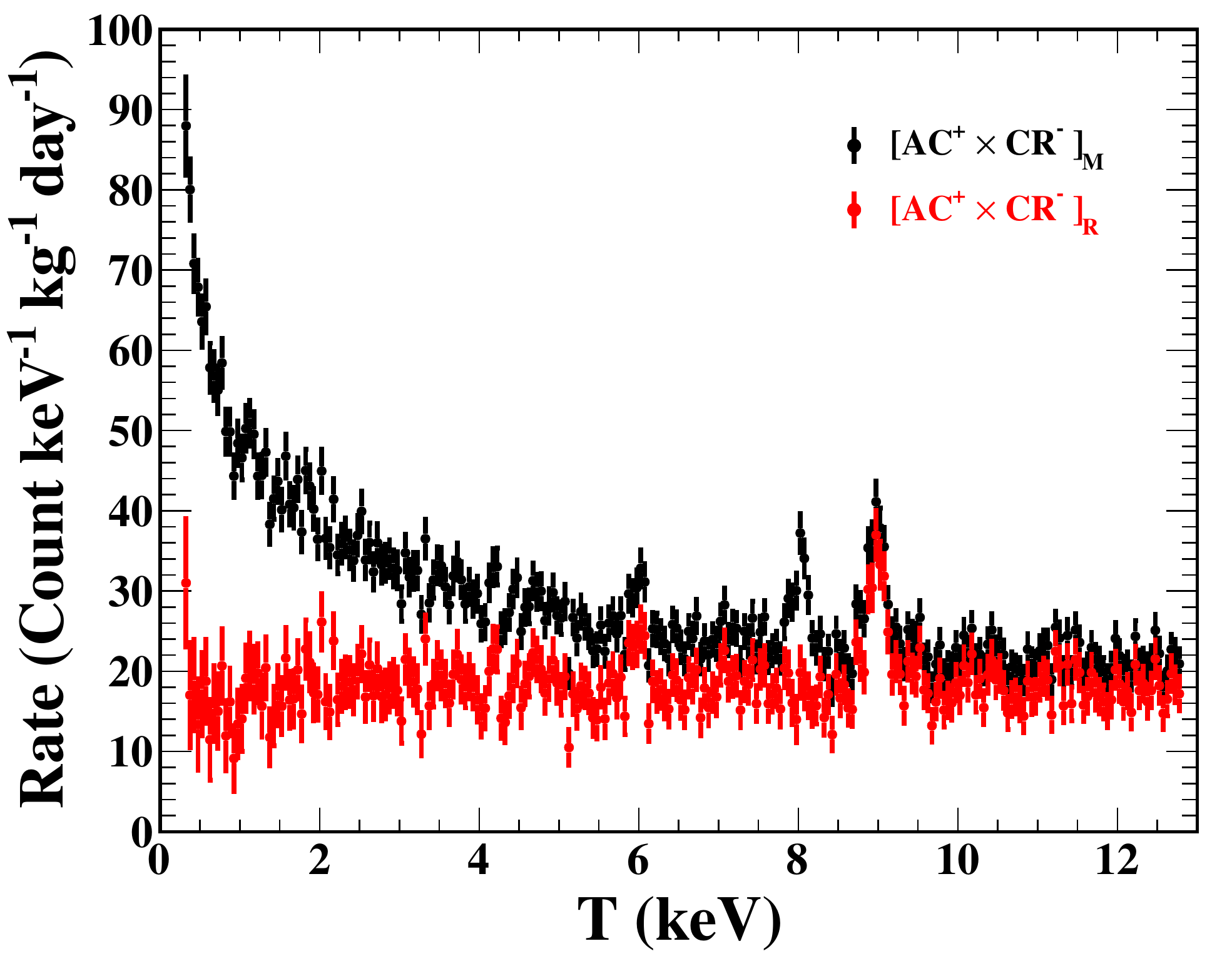}
    \caption{Measured energy spectrum of data sample,
      $\vt$, before and after cosmic ray correction.}
    \label{fig::vtcorr}
  \end{minipage}
  \hfill
\end{figure*}

Cosmic ray muon detection efficiency of CRV system on average for a week {period} 
is typically 93\%, and the rejection efficiency of cosmic ray 
induced event by timing correlation between Ge-trigger and CR-trigger is 92\%. 
The measured $\vv$ spectrum therefore could be contaminated
with cosmic ray induced events. The presence of additional cosmic 
events in the residual energy spectrum of $\vv$ 
events is suppressed by the statistical background model.
The real cosmic (T$_{r}$) and non-cosmic (V$_{r}$) event rate 
can be correlated with measured (V$_{m}$ ,T$_{m}$ ) rate via 
following couple of equations
\begin{eqnarray}
&&V_{r} = \frac{\lambda_{crv}}{\epsilon_{crv}+\lambda_{crv}-1}. V_{m} - \frac{1-\lambda_{crv}}{\epsilon_{crv}+\lambda_{crv}-1}. T_{m}~~~~~\mathrm and
\label{EQ:real-rateV} 
\end{eqnarray}
\begin{eqnarray}
&&T_{r} = \frac{\epsilon_{crv}}{\epsilon_{crv}+\lambda_{crv}-1}. T_{m} - \frac{1-\epsilon_{crv}}{\epsilon_{crv}+\lambda_{crv}-1}. V_{m},
\label{EQ:real-rateT}  
\end{eqnarray}
where $\epsilon_{crv}$ is the rejection efficiency of cosmic ray cut and
$\lambda_{crv}$ is the detection efficiency of cosmic ray system. 
The verification of the correction method is provided by the intensity of 
known K-shell internal X-rays and ambient $\gamma$-ray energy spectrum.
The intensities of K-shell X-rays increased by putting back leakage 
events into their own sample, as illustrated in Fig.~\ref{fig::vvcorr}.
High energy $\gamma$-rays from ambient radioactivity produce
flat electron-recoil background at low energy, which is verified by simulation
and measurements ($^{241}$Am and $^{137}$Cs source).
Figure~\ref{fig::vtcorr} demonstrates the flat energy spectra after correction 
for $\vt$ data sample.

\begin{figure} 
  \centering
  \includegraphics[width=9.5cm]{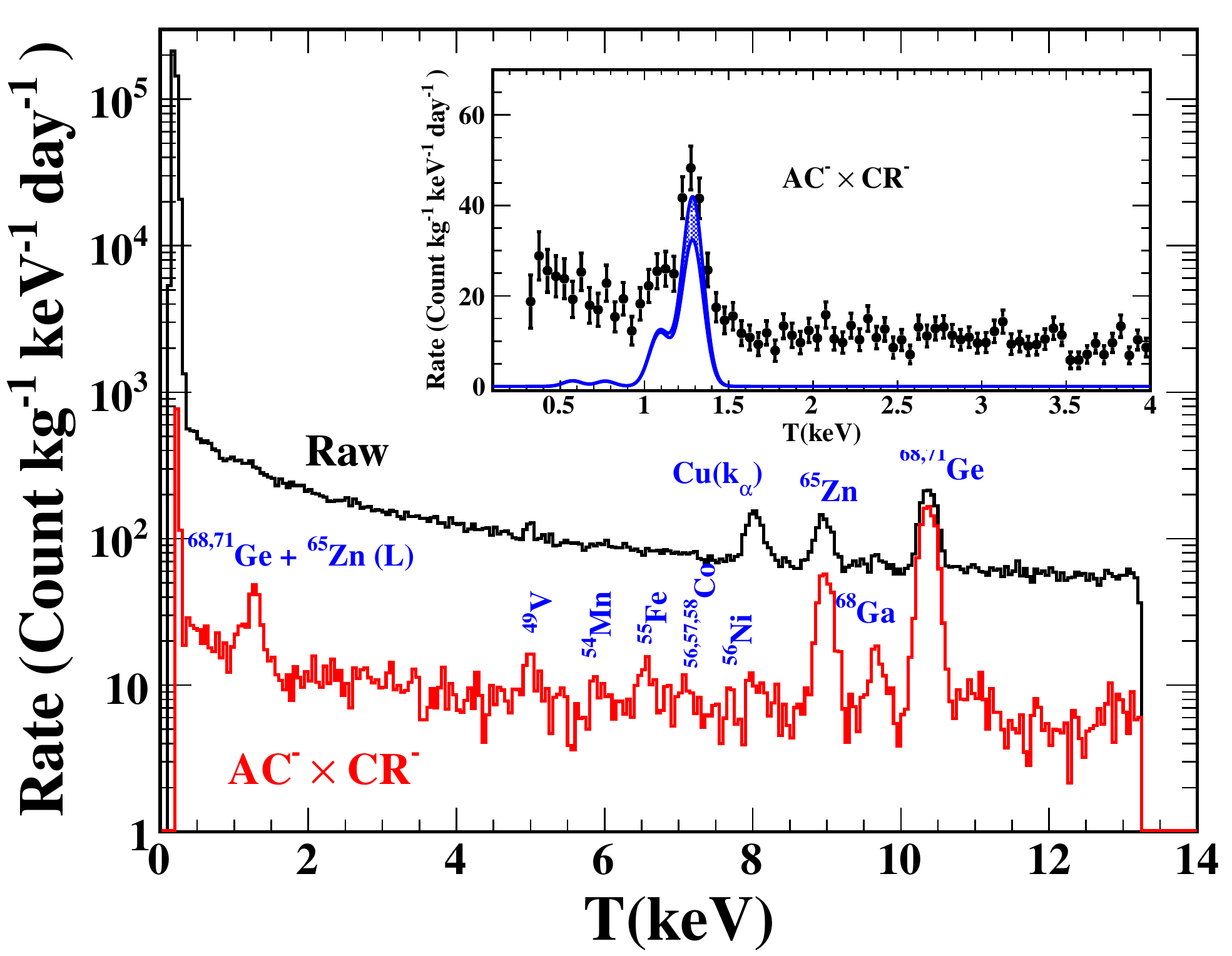}
  \caption{Measured Raw and $\vv$ spectra with $n$PCGe, the
    L-shell X-ray lines predicted from the intensities of measured K-shell X-ray lines is 
    shown in inset.}
  \label{fig::npcspectra}
\end{figure}

The observed $\vv$ spectra with internal X-ray lines from cosmogenically-induced isotopes
decays via electron capture process is shown in Fig.~\ref{fig::npcspectra}.

\begin{figure}
  \centering
  \includegraphics[width=9.5cm]{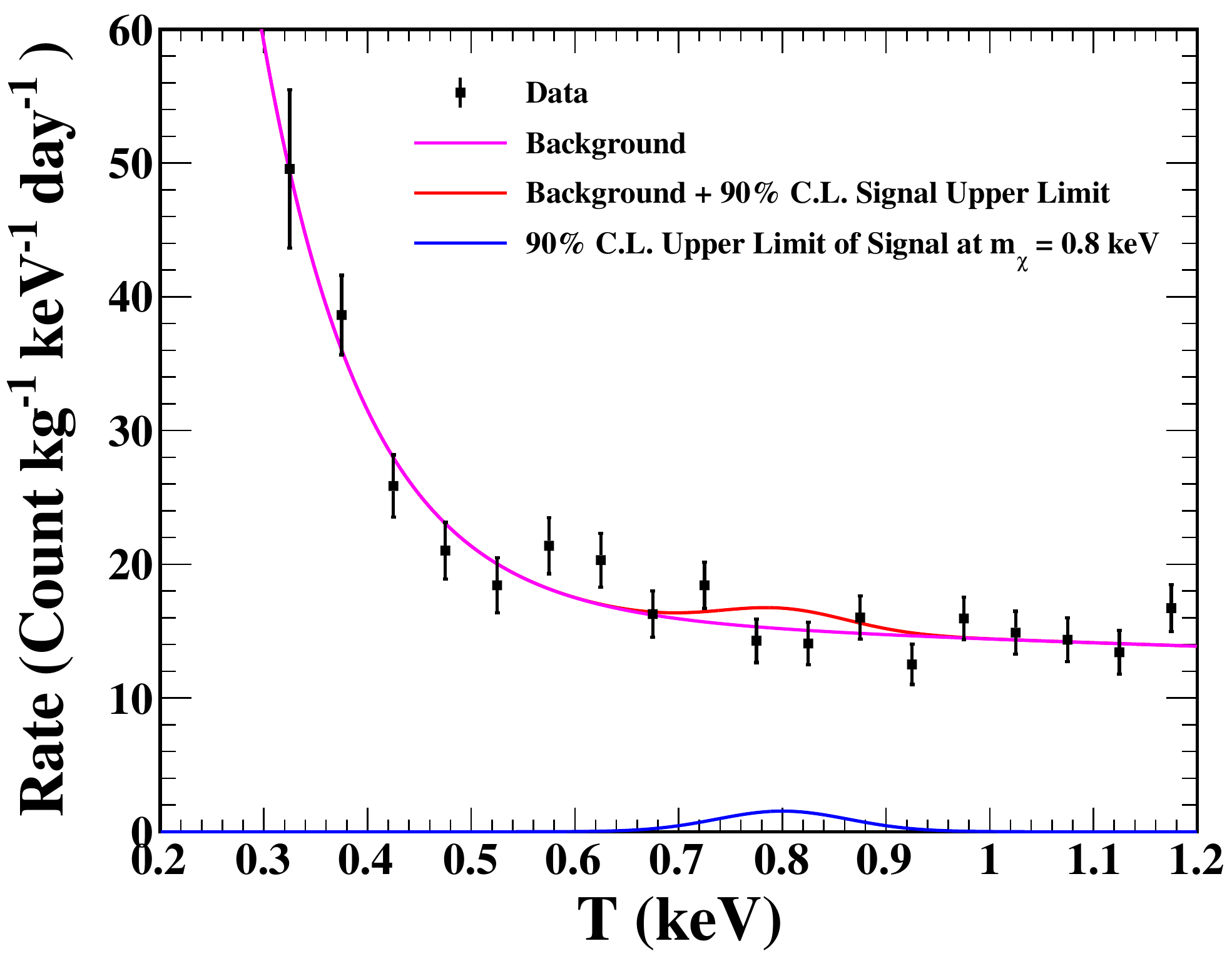}
  \caption{The efficiency-corrected energy spectrum near to the
    analysis threshold with the whole exposure. An example of excluded 
    rate (90\%~C.L.) at 0.8 keV for non-relativistic bosonic $\chi$
    signal is shown with blue curve. The assumption of flat plus exponential
    background is also superimposed.}
  \label{fig::90clsignal}
\end{figure}
The intensity of K-shell peaks decay with time and low energy 
L-capture lines corresponding to $^{65}$Zn and $^{68}$Ge behaves 
similar to high energy K-capture lines. The L/K-capture ratio 
from the same isotope has been well understood by theoretically 
and independent experimental measurements. Therefore, the 
intensities of L-shell X-ray lines are estimated 
from measured K-shell X-ray lines and subtracted from 
measured $\vv$ spectra. The predicted intensities of L-shell X-ray lines
are depicted in the inset of Fig.~\ref{fig::npcspectra}.

\section{Experimental Constraints}
\label{sec::results}
\begin{figure}
  \centering
  \includegraphics[width=8.0cm]{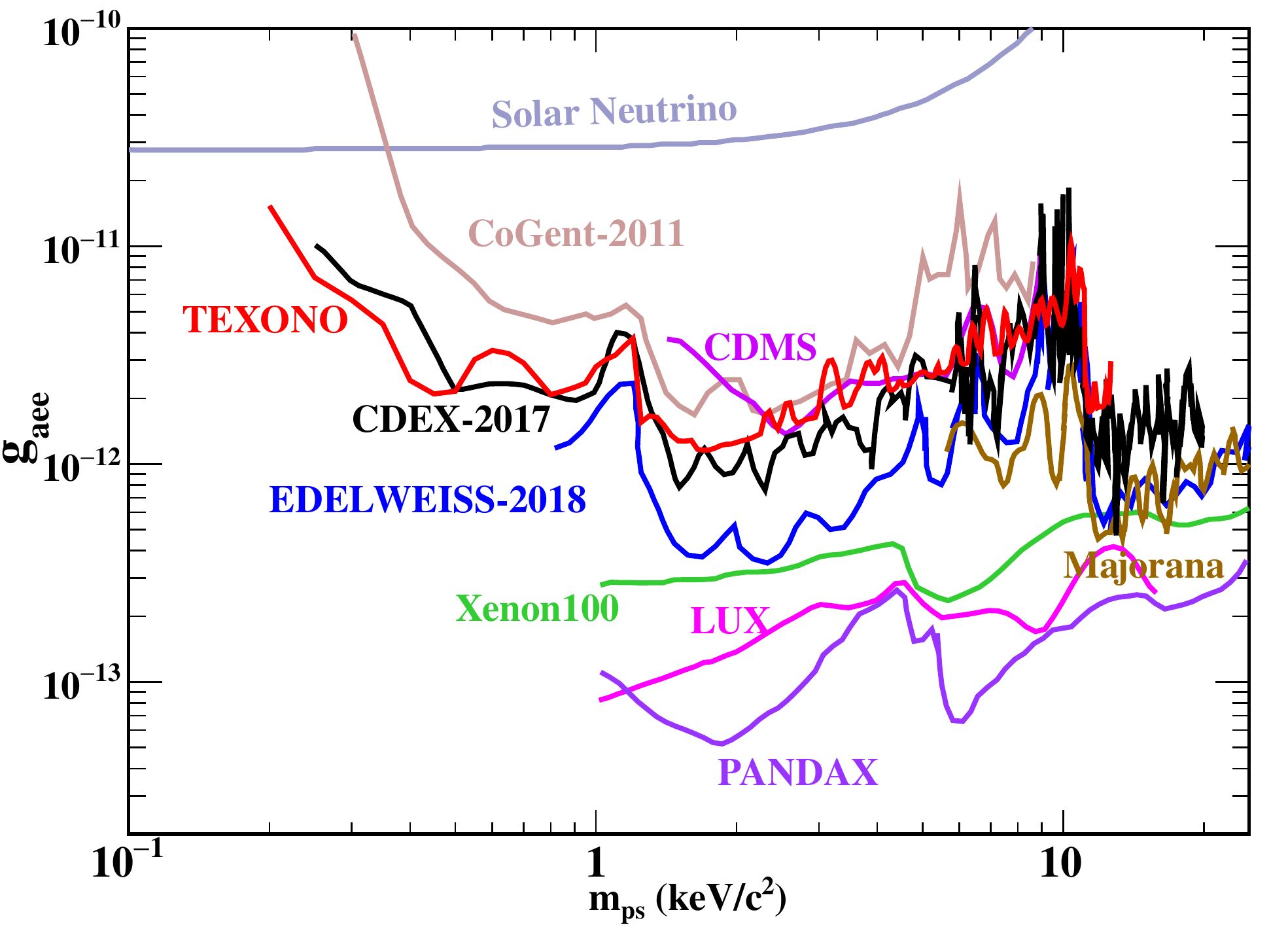}
  \caption{Constraints on $g_{aee}$ of $\chis$ as a function of mass.
    The TEXONO upper limits, at 90\%~C.L. is shown in red curve.
    Results from representative experiments CDEX~\cite{CDEX:2017}, CoGeNT~\cite{CoGeNT:2011},
    EDELWEISS~\cite{Edelw:2018}, LUX~\cite{LUX:2017}, Majorana Demonstrator~\cite{Majorana:2017},  
    PANDAX-II~\cite{PANDAX:2017}, XENON100~\cite{Xe100:2017} along with
    indirect astrophysical bound from solar neutrinos~\cite{SolarNu:2009} are also shown.}
\label{fig::ex-scalar}
\end{figure}

The signal of nonrelativistic $\chi$ would be $\delta$-function
centered at its mass, smearing by the energy resolution of the detector.
The expected signal rate of $\chi$ at measurable energy is obtained by
the convolution of interaction rate from Eq.~(\ref{eq::scalar-rate})~or~(\ref{eq::vector-rate})
with the energy resolution of the detector:
\begin{equation}
\frac{dR_{\chi}}{dE}(E,m_{\chi}) = \Phi_{\chi}(m_{\chi})\sigma_{\chi} \times \frac{1}{\sqrt{2 \pi} \sigma_{E}(m_{\chi})} e^{-\frac{(E-m_{\chi})^{2}}{2 \sigma_{E}^{2}(m_{\chi})}}   
\end{equation}
where $\chi$ = $\chis$, $\chiv$ represents pseudoscalar  and vector particles, respectively.  
A~minimal-chi-square analysis is applied with free parameters
describing a flat plus exponential background and coupling constant of $\chi$.
Our data are compatible with the background only hypothesis, and no excess of events
are observed over the background as depicted in Fig.~\ref{fig::90clsignal}.
The 90\%~confidence level~(C.L.) excluded rate in kg$^{-1}$day$^{-1}$ for a given
m$_{\chi}$ is evaluated from best fit parameters. The excluded value of coupling constant is 
then derived from excluded count rate  using  Eqs.~(\ref{eq::scalar-rate})~or~(\ref{eq::vector-rate}).
 The 90\%~C.L. exclusion limits for coupling of $\chis$ as
a function of $m_{\chi}$ are shown in Fig.~\ref{fig::ex-scalar}, results from other representative
experiments are also superimposed.

\begin{figure}
  \centering
  \includegraphics[width=8.0cm]{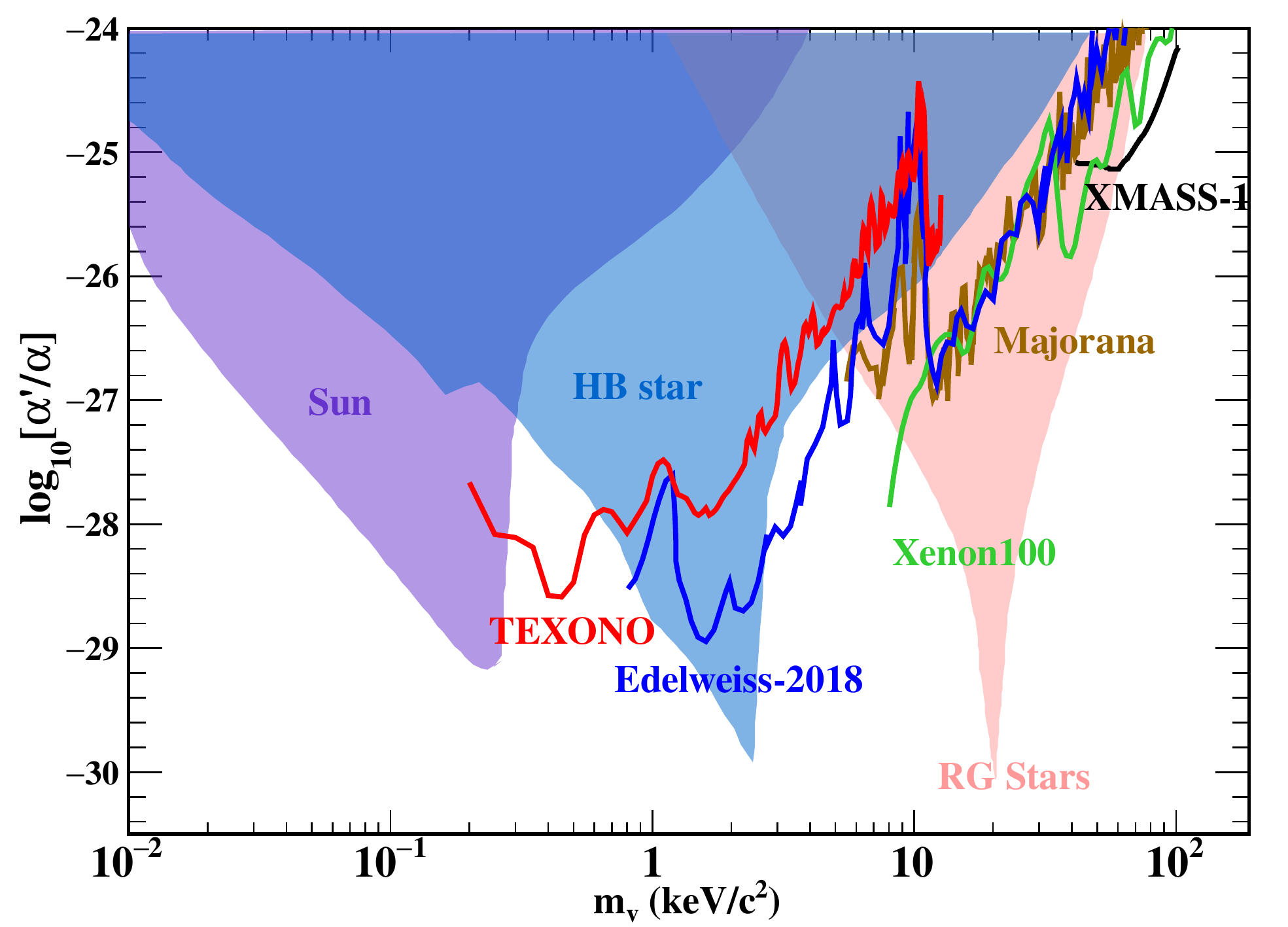}
    \caption{The 90\%~C.L. bounds on vector bosonic DM coupling from the TEXONO compared with
      astrophysical sources~\cite{DP-astro:2015} as well as experimental results from
      EDELWEISS~\cite{Edelw:2018}, Majorana Demonstrator~\cite{Majorana:2017}, XENON100~\cite{Xe100-superWIMP:2017} and XMASS-I~\cite{XMASS:2014}.}
    \label{fig::ex-vector}
\end{figure}

The upper limits on the electronic
coupling of $\chiv$ at 90\%~C.L. are derived using the same data
and analysis technique. Figure~\ref{fig::ex-vector} shows the
exclusion limits at 90\%~C.L. of electronic coupling of $\chiv$.
The laboratory upper limits on vector-electric coupling
established by direct detection experiments are potentially more
stringent than limits derived from cosmological and astronomical sources. 
The observed fluctuations in the exclusion curve correspond to
statistical fluctuations in the background. Our data with sub-keV sensitivity
provide the improved constraints on electronic couplings of pseudoscalar and vector 
bosonic DM in sub-keV/c$^{2}$ mass region. 

\section{Conclusion and prospects}
\label{sec::summary}
New constraints on dark matter candidates $\chis$ and $\chiv$ with $n$PCGe are
presented in this article. This detector technique provides low threshold, excellent
energy resolution and without complications of anomalous surface background events.
It is therefore an optimal detector to study not just $\chi$ but also milli-charged
neutrino~\cite{txn-mQ:2014} and dark matter WIMPs~\cite{CDEX10:2018}, as well as
dark matter heavy sterile neutrinos~\cite{ntuSterile:2016}. Data taking continues at
KSNL with improved PCGe detectors. The goals are to enhance the sensitivities of these
measurements, as well as to study the standard model neutrino-nucleus coherent
scattering~\cite{Soma:2014, Kerman:2016, Wong:2018}.

\section{Acknowledgments}
This work is supported by the 2017-21 Academia Sinica Investigator Award AS-IA-106-M02,
as well as contracts MOST~107-2119-M-001-028-MY3 from the Ministry of Science and
Technology, and 2017-18/ECP-2 from the National Center of Theoretical Physics, Taiwan.

\bibliography{bdm_draft}

\end{document}